\documentclass[10pt]{article}

\def\AnnENS{Ann.~\'Ec.~Norm.~}
\def\CRAS{C.~R.~Acad.~Sc.~Paris}
\def\LMP{Lett.~Math.~Phys.~}
\def \D {\hbox{d}}
\def \tr{\mathop{\rm tr}\nolimits}
\def \diag{\mathop{\rm diag}\nolimits}

\def\PVI    {{\rm P6}}
\def\qPVI   {{\rm q-P6}}
\def\abcd{\alpha,\beta,\gamma,\delta}

\begin{document}

\title{On the Lax pairs of the sixth Painlev\'e equation\footnote
{\textit{RIMS K\^oky\^uroku Bessatsu}, to appear (2007).
Kyoto, 15--20 May 2006.
Preprint S2006/074.
nlin.SI/0701049
}
}

\author{Robert Conte
{}\\
\\ \dag Service de physique de l'\'etat condens\'e (URA 2464), CEA--Saclay
\\ F--91191 Gif-sur-Yvette Cedex, France
\\ E-mail:  Robert.Conte@cea.fr
}

\maketitle

\hfill 

{\vglue -10.0 truemm}
{\vskip -10.0 truemm}

\begin{abstract}
The dependence of the sixth equation of Painlev\'e
on its four parameters
$(2 \alpha,-2 \beta,2 \gamma,1-2 \delta)
=(\theta_{\infty}^2,\theta_{0}^2,\theta_{1}^2,\theta_{x}^2)$
is holomorphic,
therefore one expects all its Lax pairs to display such a dependence.
This is indeed the case of the second order scalar ``Lax'' pair of Fuchs,
but the second order matrix Lax pair of Jimbo and Miwa
presents a meromorphic dependence on $\theta_\infty$
(and a holomorphic dependence on the three other $\theta_j$).
We analyze the reason for this feature
and make suggestions to suppress it.
\end{abstract}




\noindent \textit{PACS 1995}
  02.20.Qs   
  11.10.Lm   

\baselineskip=12truept


\tableofcontents

\section{Introduction}

Consider a second order linear ordinary differential equation
for $\psi(t)$
with five Fuchsian singula\-rities,
one of them $t=u$ being apparent
(i.e.~the ratio of two linearly independent solutions
remains single valued around it)
and the four others having a crossratio $x$.
The condition that the ratio $\psi_1/ \psi_2$ of two linearly
independent solutions be singlevalued when $t$ goes around any
of these singularities results in one constraint between $u$ and $x$,
which is \cite{FuchsP6}
that the apparent singularity $u$,
considered as a function of the crossratio $x$,
obeys the sixth Painlev\'e equation $\PVI$.
In its normalized form (choice $(\infty,0,1,x)$ of the four
nonapparent Fuchsian singularities),
this ODE is
\cite{FuchsP6} 
\begin{eqnarray*}
E(u) & \equiv &
-u''
+ \frac{1}{2} \left[\frac{1}{u} + \frac{1}{u-1} + \frac{1}{u-x} \right] u'^2
- \left[\frac{1}{x} + \frac{1}{x-1} + \frac{1}{u-x} \right] u'
\\
& &
+ \frac{u (u-1) (u-x)}{x^2 (x-1)^2}
  \left[\alpha + \beta \frac{x}{u^2} + \gamma \frac{x-1}{(u-1)^2}
        + \delta \frac{x (x-1)}{(u-x)^2} \right]
 =  0,
\end{eqnarray*}
its four parameters $\abcd$ representing
the differences $\theta_j$ of the two Fuchs indices
at the four nonapparent singularities $t=\infty,0,1,x$,
\begin{eqnarray}
& &
(2 \alpha,-2 \beta,2 \gamma,1-2 \delta)
=(\theta_{\infty}^2,\theta_{0}^2,\theta_{1}^2,\theta_{x}^2).
\label{eqLaxP6Theta}
\end{eqnarray}

The proof by Poincar\'e \cite{Poincare1883}
of the impossibility to remove the apparent
singularity in the second order scalar isomonodromic deformation
certainly motivated Jimbo and Miwa to consider,
in place of the scalar isomonodromy problem,
the matrix isomonodromy problem of the same order (two),
\begin{eqnarray}
& &
\partial_x \psi = L \psi,\quad
\partial_t \psi = M \psi,\
[\partial_x - L, \partial_t - M] = 0.
\label{matrixLax}
\end{eqnarray}
There indeed exists a choice \cite{JimboMiwaII}
of second order matrices $(L,M)$
whose isomonodromy condition also yields $\PVI$,
in which the singularities of the monodromy matrix $M$
in the $t$ complex plane
are four Fuchsian points of crossratio $x$,
without the need for an apparent singularity.

This beautiful result
however presents the drawback to have a meromorphic dependence
on one of the four monodromy exponents $\theta_j$,
while $u''$ in $\PVI$ has a holomorphic such dependence.
The purpose of this work is to explore several directions
in order to remove this drawback from matrix Lax pairs.

A possibility to achieve that is to consider some simple physical system
admitting a Lax pair and a reduction to $\PVI$.
The corresponding reduction of its Lax pair could then provide
a holomorphic Lax pair of $\PVI$.
One such system if the three-wave resonant interaction,
but the resulting Lax pair has third order,
and its reduction to second order still encounters some obstacles \cite{CGM}.
The Maxwell-Bloch system \cite{SM}
could be a better candidate because its Lax pair is second order.

The paper is organized as follows.
In section \ref{sectionGarnierPair},
we recall the scalar ``Lax'' pair of Richard Fuchs,
because its expression is required later on.

In section \ref{sectionMatrix_isomonodromy},
we point out the meromorphic dependence in the second order Lax pair
obtained by matrix monodromy.

In section \ref{sectionTowards_holomorphic_matrix_Lax},
we define in some detail the small amount of required computations 
in order to obtain a holomorphic Lax pair.

In section \ref{sectionFirst_attempt},
we explore the simplest possibility beyond the assumption of Jimbo and Miwa.
The resulting Lax pair is linked to a type studied by Kimura 
\cite{Kimura1981}
and the matrix elements are algebraic functions of $u',u,x$
while in the JM case they are rational functions.

\section{Holomorphic Lax pair by scalar isomonodromy}
\label{sectionGarnierPair}
\indent

This pair \cite{FuchsP6,Fuchs1907},
as more nicely written in Ref.~\cite{GarnierThese},
is characterized by the two homographic invariants $(S,C)$,
\begin{eqnarray}
& &
\partial_t^2 \psi + (S/2) \psi=0,\
\label{eqGarnierP6LaxODE}
\\
& &
\partial_x \psi + C \partial_t \psi -(1/2) C_t \psi = 0,\
\label{eqGarnierP6LaxPDE}
\end{eqnarray}
with the commutativity condition,
\begin{eqnarray}
& &
X \equiv S_x + C_{ttt} + C S_t + 2 C_t S=0,
\end{eqnarray}
where
\begin{eqnarray}
& &
{\hskip -10.0 truemm}
- C=\displaystyle\frac{t(t-1) (u-x)}{(t-u) x(x-1)},
  \quad
\label{eqC}
\\
& &
{\hskip -10.0 truemm}
-\frac{S}{2}=
\frac{3/4}{(t-u)^2}
+ \frac{\beta_1 u' + \beta_0}{(t-u) t(t-1)}
+\frac{[(\beta_1 u')^2 - \beta_0^2] \displaystyle\frac{u-x}{u (u-1)}
+ f_{\rm G}(u)}{t(t-1)(t-x)}
+ f_{\rm G}(t),
\label{eqS}
\\
& &
{\hskip -10.0 truemm}
\beta_1=\displaystyle-\frac{x (x-1)}{2 (u-x)},
  \quad
\beta_0=-u+\frac{1}{2},
\\
& &
{\hskip -10.0 truemm}
f_G(z)=\displaystyle\frac{a}{z^2} + \frac{b}{(z-1)^2}
\displaystyle+\frac{c}{(z-x)^2} + \frac{d}{z (z-1)},
\\
& &
{\hskip -10.0 truemm}
(2\alpha, -2\beta, 2\gamma, 1-2\delta) =(4(a+b+c+d+1),4 a+1,4 b+1,4 c+1).
\end{eqnarray}

Like $u''$ in the definition of $\PVI$,
this scalar Lax pair depends holomorphically on
the four $\theta_j$,
and also on their squares.
Its singularities in the complex plane of $t$ are 
the five Fuchsian points $t=\infty,0,1,x,u$,
among which $t=u$ is apparent.

\section{Meromorphic Lax pair by matrix isomonodromy}
\label{sectionMatrix_isomonodromy}

Let us introduce the Pauli matrices $\sigma_k$
\begin{eqnarray}
& &
\sigma_1=\pmatrix{0 & 1 \cr 1 & 0 \cr },\
\sigma_2=\pmatrix{0 &-i \cr i & 0 \cr },\
\sigma_3=\pmatrix{1 & 0 \cr 0 &-1 \cr },\
\sigma_j \sigma_k= \delta_{jk} + i \varepsilon_{jkl} \sigma_l,\
\label{eqPauli}
\\
& &
\sigma^{+}=\pmatrix{0 & 1 \cr 0 & 0 \cr },\
\sigma^{-}=\pmatrix{0 & 0 \cr 1 & 0 \cr }.
\nonumber
\end{eqnarray}

As proven in \cite{JimboMiwaII},
the apparent singularity of the scalar Lax pair 
can be removed by considering a second order matrix Lax pair,
\begin{eqnarray}
& & 
\partial_x \Psi=L \Psi,\
\partial_t \Psi=M \Psi,\
\label{eqMatrixPair}
\end{eqnarray}
and defining the monodromy matrix $M$ as the sum of four Fuchsian
singularities $t=\infty,0,1,x$,
\begin{eqnarray}
& & 
M = \frac{M_0(x)}{t} + \frac{M_1(x)}{t-1} + \frac{M_x(x)}{t-x}, \quad
M_\infty+M_0+M_1+M_x = 0.
\label{eqM}
\end{eqnarray}
However,
in order to integrate the differential system of the monodromy conditions,
\begin{eqnarray}
& & 
\forall t:\ L_t - M_x + L M - M L=0.
\label{eqMatrixIsomonodromy}
\end{eqnarray}
the choice of $L$ is not unique and
the type of dependence of $L(x,t)$ on $t$ must be an input.
With the very convenient choice \cite{JimboMiwaII} of a simple pole at the
crossratio $t=x$,
\begin{eqnarray}
& & 
L=- \frac{M_x}{t-x},\
M = \frac{M_0(x)}{t} + \frac{M_1(x)}{t-1} + \frac{M_x(x)}{t-x}, \quad
M_\infty+M_0+M_1+M_x = 0,
\label{eqMatrixChoiceJM}
\end{eqnarray}
and after minor transformations \cite{Cargese1996Mahoux}
mainly aimed at making all entries $(L_{jk},M_{jk})$ algebraic
(not only with algebraic logarithmic derivatives),
one obtains the traceless, algebraic Lax pair,
\begin{eqnarray}
L
&= &
- \frac{M_x}{t-x} + L_\infty,\
M=\frac{M_0}{t} + \frac{M_1}{t-1} + \frac{M_x}{t-x},\
\label{eqLaxP6JMTracelessAlgebraic}
\\
L_\infty
&= &
 -\frac{(\Theta_\infty-1)(u-x)}{2x(x-1)}            \sigma_3,
\\
2 M_\infty
&= &
 \Theta_\infty                                               \sigma_3,
\\
2 M_0
&= &
 z_0                                                         \sigma_3
 - \frac{u}{x}                                               \sigma^+
 +(z_0^2-\theta_0^2) \frac{x}{u}                             \sigma^-,
\\
2 M_1
&= &
 z_1 \sigma_3 + \frac{u-1}{x-1}                              \sigma^+
 -(z_1^2-\theta_1^2) \frac{x-1}{u-1}                         \sigma^-,
\\
2 M_x
&= &
 \left(
 (\theta_0^2-z_0^2)\frac{x}{u} - (\theta_1^2-z_1^2) \frac{x-1}{u-1}
 \right)                                                     \sigma^-
-\frac{u-x}{x(x-1)}                                          \sigma^+
\nonumber
\\
& &
- (\Theta_\infty+z_0+z_1)                                    \sigma_3,
\\
z_0
&= &
\frac{1}{2 \Theta_\infty x (u-1)(u-x)}
\bigg[
\left(x(x-1)u'-(u-1)(u-\Theta_\infty (u-x))\right)^2
\nonumber
\\
& &
 - (\Theta_\infty^2+\theta_0^2) x         (u-1) (u-x)
 + \theta_1^2                     (x-1) u       (u-x)
 - \theta_x^2                   x (x-1) u (u-1)
\bigg],
\nonumber
\\
z_1
&= &
\frac{-1}{2 \Theta_\infty (x-1)u(u-x)}
\bigg[
\left(x(x-1)u'-u (u-1-\Theta_\infty (u-x))\right)^2
\nonumber
\\
& &
 + (\Theta_\infty^2+\theta_1^2)   (x-1) u       (u-x)
 - \theta_0^2                   x         (u-1) (u-x)
 - \theta_x^2                   x (x-1) u (u-1)
\bigg],
\nonumber
\\
& &
(2 \alpha,-2 \beta,2 \gamma,1-2 \delta)
=((\Theta_{\infty}-1)^2,\theta_{0}^2,\theta_{1}^2,\theta_{x}^2).
\nonumber
\end{eqnarray}

The origin of the meromorphic dependence in
(\ref{eqLaxP6JMTracelessAlgebraic}),
as displayed in $z_0$ and $z_1$,
seems to be the simplifying assumption \cite{JimboMiwaII}
that the residue $M_\infty$ can be chosen diagonal,
\begin{eqnarray}
& &
M_\infty=\frac{\Theta_\infty}{2} \sigma_3.
\label{eqJMDiagonalAssumption}
\end{eqnarray}
Indeed,
when $\Theta_\infty$ vanishes,
the residue also vanishes and one singular point is lost,
thus preventing to obtain $\PVI$ which requires four
nonapparent singular points.

As an additional motivation of the present work,
this meromorphic feature is also present
in many discrete Lax pairs of discrete $\PVI$ equations,
for instance in the Lax pair found by Jimbo and Sakai \cite{JS1996},
as an output to the matrix discrete isomonodromy problem
\begin{eqnarray}
& &
Y( x, q t) = A(x,t) Y(x,t),
\\
& &
A  = A_0(x) + A_1(x) t + A_2(x) t^2,
\end{eqnarray}
where 
$x$ is the independent variable,
$t$ is the spectral parameter,
and
the matrix $A$ defines four singular points in the $t$ complex plane.
If the residue $A_2$ at $t=\infty$ is chosen diagonal
\cite[Eq.~(10)]{JS1996},
\begin{eqnarray}
& &
A_2=\diag(\kappa_1,\kappa_2),
\end{eqnarray}
then the Lax pair contains the denominator $\kappa_1 - \kappa_2$
and, when $\kappa_1 = \kappa_2$, 
the isomonodromy problem cannot yield a $\qPVI$ equation.

\section{Towards a holomorphic matrix Lax pair}
\label{sectionTowards_holomorphic_matrix_Lax}

In order to get rid of this unwanted meromorphic dependence,
let us change the assumptions on the matrix Lax pair $(L,M)$
along the lines explored in Ref.~\cite{LCM2003}.
For the assumption (\ref{eqM}) on $M$, which must be kept,
we adopt the convention
\begin{eqnarray}
& &
\tr M_j=0,\
\det M_j=-\frac{\theta_j^2}{4}=\hbox{constant},\
j=\infty,0,1,x,
\label{eqResiduesInvariants}
\end{eqnarray}
and we represent the four residues so as to preserve the invariance 
under permutation,
\begin{eqnarray}
& &
M_j=\frac{1}{2}
\pmatrix{z_j & (\theta_j-z_j) u_j \cr (\theta_j+z_j) u_j^{-1} & -z_j \cr },\
j=\infty,0,1,x,
\label{eqResiduesAssumption}
\end{eqnarray}
in which $z_j,u_j$ are functions of $x$.

After an assumption has been chosen for the dependence of $L(x,t)$ on $t$,
there is no need to integrate the monodromy conditions 
(\ref{eqMatrixIsomonodromy}).
Indeed,
one \textit{a priori} knows
that their general solution is expressed in terms of a $\PVI$ function.
Therefore a ``lazy'' method to perform the integration
is to first convert the matrix Lax pair
(\ref{eqMatrixPair}) to scalar form,
then to identify the result with the scalar Lax pair
(\ref{eqGarnierP6LaxODE})--(\ref{eqGarnierP6LaxPDE}).

Let us denote
$\Psi={}^{\rm t}(\psi_1\ \psi_2)$
the base vectors of
the matrix Lax pair after rotation by an arbitrary constant angle $\varphi$,
\begin{eqnarray}
& & 
P=\pmatrix{\cos \varphi & \sin \varphi \cr -\sin \varphi & \cos \varphi \cr},\
\partial_x \Psi=P^{-1} L P \Psi,\
\partial_t \Psi=P^{-1} M P \Psi.
\label{eqRotation}
\end{eqnarray}
After elimination of $\psi_2$
and removal of the first derivative $\psi_1'$ in the resulting
second order linear ODE for $\psi_1$,
the identification of the two sets of coefficients $(S,C)$
will provide $L$ and $M$ in terms of a solution $u$ of $\PVI$.

Whatever be the assumption for $L$,
the three scalar conditions of zero sum for the residues,
\begin{eqnarray}
& &
M_\infty+M_0+M_1+M_x = 0,
\label{eqZeroSumRes}
\end{eqnarray}
under the condition that $u_0,u_1,u_x$ are all different,
are first solved for $z_0,z_1,z_x$,
\begin{eqnarray}
& & 
\left\lbrace
\begin{array}{ll}
\displaystyle{
\frac{J}{u_1-u_x} z_0=
 z_\infty \left(u_1^{-1} + u_x^{-1}\right)
-(\theta_\infty - z_\infty) u_\infty u_1^{-1} u_x^{-1}
-(\theta_\infty + z_\infty) u_\infty^{-1}
}\\ \displaystyle{\phantom{0123456789}
+\theta_0 \left(u_0 u_1^{-1} u_x^{-1} - u_0^{-1}\right)
+\theta_1 \left(             u_x^{-1} - u_1^{-1}\right)
+\theta_x \left(             u_1^{-1} - u_x^{-1}\right),
}
\\
\displaystyle{
\frac{J}{u_x-u_0} z_1=
 z_\infty \left(u_x^{-1} + u_0^{-1}\right)
+(\theta_\infty - z_\infty) u_\infty u_x^{-1} u_0^{-1}
-(\theta_\infty + z_\infty) u_\infty^{-1}
}\\ \displaystyle{\phantom{0123456789}
+\theta_1 \left(u_1 u_x^{-1} u_0^{-1} - u_1^{-1}\right)
+\theta_x \left(             u_0^{-1} - u_x^{-1}\right)
+\theta_0 \left(             u_x^{-1} - u_0^{-1}\right),
}
\\
\displaystyle{
\frac{J}{u_0-u_1} z_x=
 z_\infty \left(u_0^{-1} + u_1^{-1}\right)
+(\theta_\infty - z_\infty) u_\infty u_0^{-1} u_1^{-1}
-(\theta_\infty + z_\infty) u_\infty^{-1}
}\\ \displaystyle{\phantom{0123456789}
+\theta_x \left(u_x u_0^{-1} u_1^{-1} - u_x^{-1}\right)
+\theta_0 \left(             u_1^{-1} - u_0^{-1}\right)
+\theta_1 \left(             u_0^{-1} - u_1^{-1}\right),
}
\end{array}
\right.
\label{eqSystemzjSol}
\end{eqnarray}
in which $J$ denotes the Jacobian
\begin{eqnarray}
& &
J \equiv \frac{D(M_{\infty,11},M_{\infty,12},M_{\infty,21})}{D(z_0,z_1,z_x)}
=-\frac{(u_0-u_1)(u_1-u_x)(u_x-u_0)}{u_0 u_1 u_x}.
\end{eqnarray}

\section{A Kimura-type Lax pair}
\label{sectionFirst_attempt}

Following (\ref{eqLaxP6JMTracelessAlgebraic})
and \cite[Eq.~(4.18)]{LCM2003},
let us assume
\begin{eqnarray}
& & 
L=- \frac{M_x}{t-x} + L_\infty,\
L_\infty= m(x) M_\infty,
\label{eqLaxP6_Ltry1}
\end{eqnarray}
which defines the differential system
\begin{eqnarray}
& & 
\left\lbrace
\begin{array}{ll}
\displaystyle{
M_0'=\frac{\lbrack M_x,M_0\rbrack}{x}  - m \lbrack M_\infty,M_0\rbrack,\
}\\ \displaystyle{
M_1'=\frac{\lbrack M_x,M_1\rbrack}{x-1}- m \lbrack M_\infty,M_1\rbrack,\
}
\\
\displaystyle{
M_x'=-\frac{\lbrack M_x,M_0\rbrack}{x}-\frac{\lbrack M_x,M_1\rbrack}{x-1}
     - m \lbrack M_\infty,M_x\rbrack.
}
\end{array}
\right.
\label{eqZeroCurvatureCond1}
\end{eqnarray}
Such a choice ensures that $M_0+M_1+M_x$ is a first integral,
and therefore $M_\infty$ a constant.
The system (\ref{eqZeroCurvatureCond1}) is equivalent to
\begin{eqnarray}
& &
z_j'=\frac{P_j(u_k,z_k,\theta_k,m)}{x(x-1)u_0 u_1 u_x},\
u_j'=\frac{Q_j(u_k,z_k,\theta_k,m)}{x(x-1)u_0 u_1 u_x},\
j \in \lbrace        0,1,x\rbrace,\
k \in \lbrace \infty,0,1,x\rbrace,\
\label{eqzj'uj'Sol}
\end{eqnarray}
in which $P_j,Q_j$ denote polynomials of their arguments,
and
the closure conditions $z_j'=(z_j)'$
between the systems (\ref{eqzj'uj'Sol})
and
(\ref{eqSystemzjSol})
are identically satisfied.


The identification of the two $C$'s of the two scalar Lax pairs
of the type
(\ref{eqGarnierP6LaxODE})--(\ref{eqGarnierP6LaxPDE})
is equivalent to the two relations
\begin{eqnarray}
& &
\left\lbrace
\begin{array}{ll}
\displaystyle{
\left\lbrack m+\frac{u-x}{x(x-1)}\right\rbrack
\left\lbrack
z_\infty-\theta_\infty
\frac{(\cos 2 \varphi+1) u_\infty+(\cos 2 \varphi-1) u_\infty^{-1}}
     {(\cos 2 \varphi+1) u_\infty-(\cos 2 \varphi-1) u_\infty^{-1}
      -2 \sin 2 \varphi}
\right\rbrack
=0,
}\\ \displaystyle{
\hbox{when } \varphi=0:\ 
 (z_\infty-\theta_\infty)u_\infty u(u-x)
+(z_0     -\theta_0     )u_0       (u-x)
-(z_x     -\theta_x     )u_x      (x-1) u
=0,
}
\end{array}
\right.
\label{eqC=C}
\end{eqnarray}
in which, for brevity, the rotation angle $\varphi$ has been set to $0$
in the second relation.

Solving the first equation in (\ref{eqC=C}) for the second factor
would result in the vanishing of $M_\infty$ with $\theta_\infty$,
hence in the same singularity of the Lax pair at $\theta_\infty=0$
than in (\ref{eqLaxP6JMTracelessAlgebraic}).
Therefore this first equation is solved for $m$,
and in the second one can eliminate $z_0,z_1,z_x$ with
(\ref{eqSystemzjSol}),
\begin{eqnarray}
& &
\left\lbrace
\begin{array}{ll}
\displaystyle{
m=-\frac{u-x}{x(x-1)},
}\\ \displaystyle{
F(z_\infty,u_\infty,\theta_\infty,u_0,\theta_0,u_x,\theta_x,u,x,
e^{i \varphi})=0,
}
\end{array}
\right.
\label{eqC=Ca}
\end{eqnarray}
in which $F$ is a polynomial of its arguments, 
of degree two in $u$ and each $u_j$.

Before transformation to the normalized form (\ref{eqGarnierP6LaxODE}),
the second order ODE for $\psi_1$ is then
\begin{eqnarray}
& & 
 (t-u) p_1(t) \frac{\D^2 \psi_1}{\D t^2}
+\frac{p_4(t)}{t(t-1)(t-x)} \frac{\D \psi_1}{\D t}
+\frac{p_6(t)}{\left\lbrack t(t-1)(t-x)\right\rbrack^2} \psi_1=0,
\label{eqpsi1ODETry1}
\end{eqnarray}
in which $p_j$ denotes polynomials of degree $j$
whose dependence on $x$ has been omitted.
The condition that $(t-u) p_1,p_4,p_6$ have a common zero $t$
(otherwise there would be two apparent singularities)
results in (when $\varphi=0$),
\begin{eqnarray}
& & 
 (z_\infty-\theta_\infty) u_\infty
=(z_0-\theta_0) u_0 \frac{x}{u^2}
=(z_1-\theta_1) u_1 \frac{x-1}{(u-1)^2}
=(z_x-\theta_x) u_x \frac{x(1-x)}{(u-x)^2},
\end{eqnarray}
and these relations imply $p_1(t)=t-u$ and
a multiplicity two for the zero $t=u$ of the element $M_{12}$ 
of the monodromy matrix $M$.
As proven in \cite{Kimura1982},
this results in a difference of $3$ between the two Fuchs indices
at the apparent singularity $t=u$,
not $2$ like in (\ref{eqS}).
The Schwarzian associated to (\ref{eqpsi1ODETry1}),
which cannot be identified to (\ref{eqS}),
must then be identified to the Schwarzian of the equation
labelled ${\rm L}_{\rm VI}^n$ in \cite{Kimura1981}.
The resulting matrix Lax pair will probably be holomorphic in the four
$\theta_j$ but
surely not rational in $u,u',x$,
since the transformation between the apparent singularities of
${\rm L}_{\rm VI}^n$  and (\ref{eqGarnierP6LaxODE})
is not birational \cite{Kimura1982}.
Therefore its explicit expression will not be given.

\section{Conclusion}

In order to build a second order matrix Lax pair of $\PVI(u,x)$
at the same time holomorphic in $\theta_j$
and rational in $u(x),u'(x),x$,
it is necessary to make an assumption for $L$ which is different
from (\ref{eqLaxP6_Ltry1}),
probably by adding to $L$ a term linear in $t$
like in \cite[\S 6]{Kimura1981} and \cite[Eq.~(4.18)]{LCM2003}.
This will be the subject of future research.

\section*{Acknowledgments}

RC thanks the Japanese organisers for their warm hospitality.


\vfill\eject
\end{document}